# Changes in magnetic scattering anisotropy at a ferromagnetic/superconducting interface.

K. Eid, H. Kurt, W.P.Pratt Jr., and J. Bass
Department of Physics and Astronomy, Center for Fundamental Materials Research, and Center for Sensor Materials, Michigan State University, East Lansing, MI 48824-2320

Abstract

We show that some metals and alloys (X = Cu, Ag, FeMn, or Cu and Ag combined with each other), sputtered between ferromagnetic Co and superconducting Nb, produce no change in current-perpendicular-to-plane magnetoresistance (CPP-MR) in a carefully designed CPP-spin-valve. In contrast, other metals (Ru or Au) or combinations (Cu or Ag combined with Au, Ru, or FeMn) change the CPP-MR, in some cases even reversing its sign. We ascribe these changes to activation of magnetic scattering anisotropies at a ferromagnetic/superconducting interface, apparently by strong spin-flipping between the Co and Nb layers.

There is now great interest in both static and transport properties of Ferromagnetic/Superconducting (F/S) metallic interfaces. Examples of topics of interest include: reductions in the superconducting transition temperature $T_c$ upon injection of a polarized current;[1] propagation of a polarized current through a superconductor;[2] proximity effects between S and F metals;[3,4] the F/S interface resistance;[5] predictions that the current-perpendicular-to-plane (CPP) magnetoresistance (MR) should be zero when measured with superconducting leads;[6] and subsequent arguments that it won't be if strong spin-flipping is present,[7] or if exchange splitting between spin up and spin down electron bands plays an important role.[8]

In this Letter we present evidence of a new phenomenon at F/S interfaces at 4.2K, activation of the magnetic scattering anisotropy at Co/Nb interfaces when different non-magnetic metals X are inserted between the Co and Nb as part of a carefully designed CPP-MR spin-valve (SV). More precisely, the SV is chosen so that when X is absent, the system SV/Nb produces almost zero change in specific resistance, $A\Delta R = AR(AP) - AR(P)$, between the parallel (P) and anti-parallel (AP) orderings of the two F-layers in the SV. Here A is the cross-sectional area through which the CPP current flows. Such a device is, thus, very sensitive to small changes in the magnetotransport properties caused by inserting X. We show that inserting X = some metals or combinations of metals leaves $A\Delta R$ unchanged, which we interpret as leaving the scattering asymmetry at the F/X interface, $\gamma_{F/X}$, approximately 0, while insertion of others (or other combinations) changes $A\Delta R$ (i.e., giving non-zero $\gamma_{F/X}$), including in some cases even its sign. We show that these changes in $A\Delta R$ correlate with the presence of strong spin-flipping between F and S, for example by showing that two metals where the anisotropy is activated, Au and Ru, flip spins much more strongly when in contact with non-superconducting Nb than does one, Cu, where the anisotropy isn't activated.

In the standard models[9-11] of CPP-MR in F/N multilayers, where N is a non-ferromagnetic metal, the F/N interfaces are characterized by two parameters. One choice is $AR_{F/N}\uparrow$ and $AR_{F/N}\downarrow$, where $\uparrow$ and $\downarrow$ means that the magnetic moment of the current-carrying-electron points along or opposite to the magnetization of the F-layer. A choice more convenient for use in the two-current series-resistor (2CSR) model[9-11] or the Valet-Fert generalization to finite spin-diffusion lengths,[10,11] is $2AR^*_{F/N} = (AR_{F/N}\uparrow + AR_{F/N}\downarrow)/2$ and $\gamma_{F/N} = (AR_{F/N}\uparrow - AR_{F/N}\downarrow)/(AR_{F/N}\uparrow + AR_{F/N}\downarrow)$. In this Letter we focus upon $\gamma_{F/N}$.

To achieve a uniform measuring current through a spin-valve of macroscopic area $A \sim 1.2$ mm$^2$, we sandwich it between crossed superconducting (S) Nb leads. Our sample geometry, preparation, and measuring techniques, are described in detail elsewhere.[12]

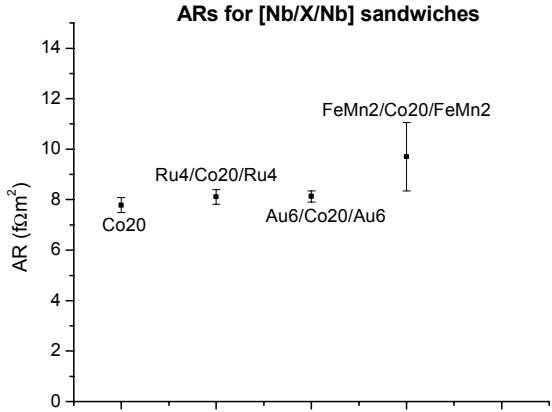

Fig. 1. ARs for Nb/X/Nb Sandwiches with X = Co(20); Ru(4)/Co(20)/Ru(4); Au(6)/Co(20)/Au(6); and FeMn(2)/Co(20)/FeMn(2). Thicknesses are in nm.

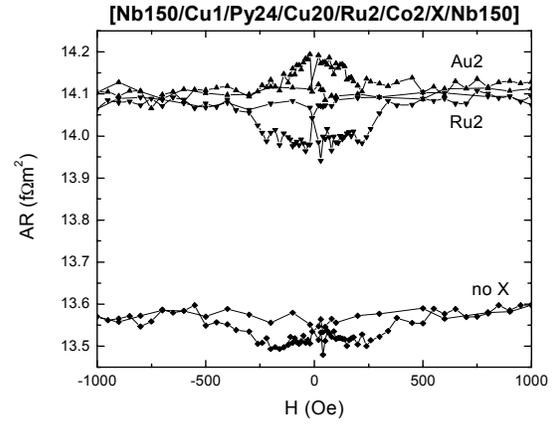

Fig. 2. AR vs H for Nb(150)/Cu(10)/Py(24)/Cu(20)/Ru(2)/Co(2)/X/Nb(150) with no X, X = Ru(2), or X = Au(2). Thicknesses are in nm.

Early experiments[9] with a 2CSR model showed that the F/S interfaces between the Nb-leads and a Co/Ag multilayer contributed only a constant term to each current channel, with interfacial spin-dependent scattering parameter $\gamma_{F/S} = 0$. We have since found[11,13,14] similar behaviors for Nb with the F-metals Fe, Ni, and the F-based alloys Permalloy (Py = $Ni_{1-x}Fe_x$ with x ≈ 0.2), and $Co_{0.91}Fe_{0.09}$. Moreover, inserting 10 nm of Cu or Ag between the F-metal Co and the S-metal Nb appeared to leave $\gamma \approx 0$,[11,13] quite different from the values[11] of $\gamma_{Co/Cu} \approx \gamma_{Co/Ag} \approx 0.8$ for Co/Cu and Co/Ag interfaces not in contact with a superconductor. Direct measurements for simple Nb/F/Nb and Nb/N/F/N/Nb sandwiches with different thicknesses of F showed that N = Cu or Ag inserts produced no systematic changes in AR.[15] As further examples, Fig. 1 shows AR for 20 nm of Co alone, and with inserts of N = 4 nm of Ru, 6 nm of Au, or 2 nm of FeMn. 20 nm of Co is thin enough so that the data all lie near the extrapolated value for zero Co thickness. We attribute the absence of increases in AR in Fig. 1 upon addition of N to at least a partial proximity effect in N (i.e. Cooper pairs propagate from S to the Co). For Cu, Ag, and Au, the residual resistivities, $\rho_o$, are too small (≤ 2 x $10^{-8}$ $\Omega m$) to noticeably increase AR. But non-superconducting FeMn ($\rho_o \sim 850$ n$\Omega m$)[16] and Ru ($\rho_o \sim 10$ n$\Omega m$)[17] should have increased AR by ~ 3.5 and 0.8 f$\Omega m^2$, respectively, each more than twice the actual deviation from the value for Co(20). FeMn is also unique in the present study, in that a layer only 1 nm thick generates very strong spin-flipping,[16] and that various of our unpublished studies lead us to believe that the FeMn/X interface is magnetically inactive (i.e., has $\gamma_{FeMn/X} = 0$).

To isolate the behavior of a single Co/X/Nb structure, we constructed a spin-valve for which A$\Delta$R is near zero when X is absent. We then examined the effect of inserting X. We achieved the required AP and P states by choosing F-metals and thicknesses to give very different saturation fields, Py(24) ($H_s \sim 20$ Oe) and Co(2) ($H_s \sim 300$ Oe). Here, and hereafter, all thicknesses are in nm. We chose a spin-valve of the form Nb/Py(24)/Cu(10)/Ru(2)/Co(2)/X/Nb. Significant spin-flipping in the Py (Py = Permalloy = $Ni_{84}Fe_{16}$) layer[14] and at Cu/Ru[17] and Co/Cu[18] interfaces means that the specific resistances of such a spin-valve cannot be described using a simple two-current series-resistor (2CSR) model. However, a modified 2CSR model so greatly simplifies the explanation of our experiment, that we use it, subject to the caveat that our explanation is only schematic, strengthened by our belief that the model does not distort the essential features of the argument. If we assume that the Py/Nb and Co/Nb interfaces are magnetically inactive (i.e. that they have $\gamma = 0$), then for no X, the 2CSR model taking account only of strong spin-flipping within the Py would give:[9-11]

$$A\Delta R \propto (\beta_{Py} \rho^*_{Py} l^{Py}_{sf} + \gamma_{Py/Cu} AR^*_{Py/Cu})(\beta_{Co} \rho^*_{Co} t_{Co} + \gamma_{Co/Ru} AR^*_{Ru/Co}), \quad (1a)$$

where $\beta_F$ is the bulk scattering anisotropy in metal F, $\rho^*_F = (\rho^{\uparrow}_F + \rho^{\downarrow}_F)/4$, and $l^{Py}_{sf}$ is the spin-diffusion length in Py.[12] The essential features of Eq. 1a are: (1) that the contributions from the Py and the Co appear separately as a product; and (2) because $\gamma_{Co/Ru}$ is negative,[17] the two terms in the second parentheses have opposite signs. Choosing $t_{Co} = 2$ nm should bring the sum of those two terms close to zero.[11,17] Fig. 2 shows that the resulting AΔR ~ − 0.05 fΩm$^2$ with no X is indeed much smaller than the value AΔR ~ + 0.4 fΩm$^2$ that we find[19] for X = Cu when there is no Ru to the left of the Co(2) layer in the spin-valve.

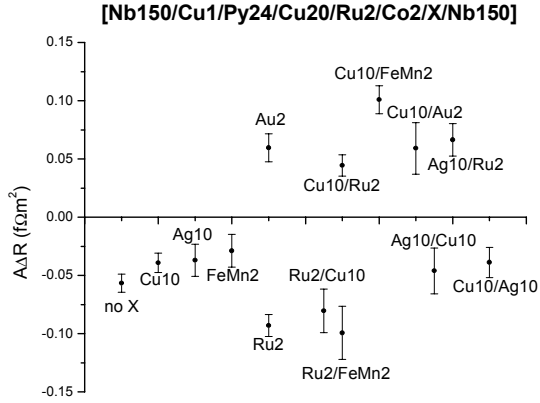

Fig. 3. AΔR for samples of the form Nb(150)/Cu(10)/Py(24)/Cu(20)/Ru(2)/Co(2)/X/Nb(150) for a variety of X, including no X at all. Thicknesses are in nm.

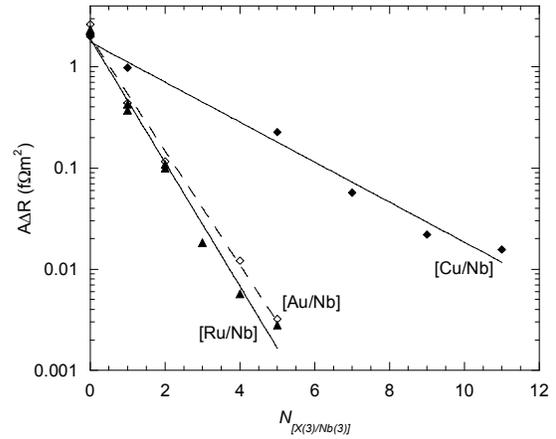

Fig. 4. AΔR vs bilayer number $N$ for inserts X = $[Ru(3)/Nb(3)]_N$, $[Au(3)/Nb(3)]_N$, and $[Cu(3)/Nb(3)]_N$ in exchange-biased spin-valves of the form Nb(150)/Cu(10)/FeMn/(8)Py(24)/Cu(10)/X/Cu(10)/Py(24)/Cu(10)/Nb(150). Thicknesses are in nm.

If, now, we insert into Eq. 1 a layer X as listed above, and if the Co/X interface is magnetically active, ($\gamma_{Co/X} \neq 0$), then this interface contributes an additional term to Eq. 1a, giving

$$A\Delta R \propto (\beta_{Py} \rho^*_{Py} l^{Py}_{sf} + \gamma_{Py/Cu} AR^*_{Py/Cu})(\beta_{Co} \rho^*_{Co} t_{Co} + \gamma_{Co/Ru} AR^*_{Co/Ru} + \gamma_{Co/X} AR^*_{Co/X}). \quad (1b)$$

If this additional term is positive ($\gamma_{Co/X} > 0$), AΔR should become more positive and if it is negative ($\gamma_{Co/X} < 0$), AΔR should become more negative. Fig. 2 shows how AΔR became more negative upon insertion of X = 2 nm of Ru, but larger and positive upon insertion of X = 2 nm of Au.

In Fig. 3 we collect together AΔR data for a series of different metals and combinations of metals. The uncertainty bars for each X were obtained by averaging the squares of the individual uncertainties in AΔR, taking the square root, and dividing by the number of such samples measured. Small uncertainties indicate several samples measured; large ones only a few. One or two cases of multiple layers involve only a single sample. To within measuring uncertainties, X = Cu, Ag, FeMn, Ag/Cu, and Cu/Ag, all leave AΔR unchanged. In contrast, X = Ru, Ru/Cu, and Ru/FeMn all make AΔR more negative, and X = Au, Cu/Ru, Cu/FeMn, Cu/Au, and Ag/Ru all invert the sign of AΔR, making AΔR positive. To within experimental uncertainties, the positive values for Au, Cu/Ru, Cu/Au, and Ag/Ru, are all the same, but that for Cu/FeMn is somewhat larger. It is of key importance to note that AΔR becomes more negative whenever Ru is next to the Co, and more positive when Cu or Ag is next to the Co but separated from the Nb by Ru, Au, or FeMn. These behaviors show that the change in AΔR is due to a change in the Co/X interface, since $\gamma_{Co/Ru} < 0$, but $\gamma_{Co/Cu} > 0$ and $\gamma_{Co/Ag} > 0$. We consider next why these differences in behavior occur. We suggest that their source might be spin-flipping (spin-memory-loss) between the Co and the Nb.

Although, as noted in the introduction, there are questions about straightforward applicability of the Taddei et al. argument[7] that the presence of strong spin-flipping is necessary for the appearance of a full CPP-MR, no other simple potential explanation for the different behaviors shown in Fig. 2 presents itself to us. We, thus, ask whether our results can be understood in terms of different amounts of spin-flipping between the Co and Nb when X is present.

All of the metallic layers N except FeMn are much thinner than their respective bulk spin-diffusion lengths. Thus, spin-flipping within those metals cannot be the source of the differences in their activities. The only potential source is spin-flipping at the N/S interface. Unfortunately, we do not have a way to measure spin-memory loss at the interfaces of these metals with superconducting Nb. We can, however, measure such loss at interfaces with non-superconducting Nb.[16] Since Au and Ru have their magnetic anisotropies activated when in contact with superconducting Nb, whereas Cu does not, we examined spin-memory loss at the interfaces of these three metals with non-superconducting Nb, using the method described in ref. [16]. In that method, a multilayer of the form [N(3nm)/Nb(3nm)]$_N$, where $N$ is the number of bilayers, is inserted into the middle of a Py-based spin-valve, in which one Py-layer is exchange-bias pinned and the other is left free to reverse in a small magnetic field. Insertion of the multilayer should then cause AΔR to decrease exponentially with $N$ as

$$A\Delta R \propto \exp(-2N\delta_{N/Nb}), \qquad (2)$$

aside from a small correction for spin-memory-loss within the $N$ layers of N and Nb. Here $\delta_{N/Nb}$ characterizes the probability of spin-flipping at the N/Nb interface. Fig. 4 shows log AΔR vs $N$ for N = Cu, Au, or Ru. The corrections for spin-memory loss within N and Nb are modest and similar for the three cases. Thus, spin-flipping at Au/Nb and Ru/Nb interfaces is several times stronger than at Cu/Nb. Spin-flipping at a Cu/Ru interface[17] is also almost as strong as at the Ru/Nb interface, providing additional help to Ru to activate the Co/Cu interface. In contrast, that at Cu/Au[20] is weaker than at Cu/Nb, leaving the entire burden for Au on the Au/Nb interface.

Turning, lastly, to FeMn, if $\gamma_{Co/FeMn} = 0$ as noted above, then strong spin-flipping in the FeMn can only activate Co/X if X = N/FeMn, where FeMn is inserted between the superconducting Nb and N. Indeed, we see in Fig. 3 that the Co/FeMn interface is not activated, but the Co/Cu/FeMn interface is.

To summarize, we constructed a multilayer that let us measure anisotropy in spin-dependent scattering at a single ferromagnetic/non-ferromagnetic metal (F/N) interface, when the other side of N is in contact with superconducting (S) Nb. We assume that the proximity effect allows Cooper pairs to pass through N, so that we are still studying an effectively F/S interface. Our results confirm earlier conclusions[11] that such an F/N interface is magnetically inactive (its insertion produces little or no change in AΔR) when N = Cu, Ag, or FeMn alone, or combinations involving only Ag and Cu. In contrast, when N = Ru or Au, or combinations involving either of these two metals or FeMn that itself is not in contact with the Co, the F/N interface becomes magnetically active, producing significant changes in AΔR including, in some cases, changes in sign. For the nominally pure metals, we showed that these magnetic activities correlate with the strength of spin-flipping at the interfaces of N with non-superconducting Nb. Qualitatively these results are consistent with the arguments of Taddei et al.[6,7] that spin-flipping can affect the CPP-MR when the sample has superconducting leads. Whether or not this argument provides the correct explanation for our data, needs further theoretical analysis.

Acknowledgments: This research was supported in part by the MSU CFMR, CSM, NSF grants DMR 02-02476 and 98-09688, and Seagate Technology.